\documentclass[12pt]{iopart}

\usepackage{graphicx}

\usepackage{iopams}

\newtheorem{theorem}{Theorem}

\newtheorem{lemma}{Lemma}

\newtheorem{proposition}{Proposition}

\newenvironment{proof}[1][Proof]{\textbf{#1.} }{\ \rule{0.5em}{0.5em}}

\begin{document}

\title[]{Casimir operators induced by Maurer-Cartan equations}

\author{Rutwig Campoamor-Stursberg\dag}

\address{\dag\ Dpto. Geometr\'{\i}a y Topolog\'{\i}a\\Fac. CC. Matem\'aticas\\
Universidad Complutense de Madrid\\Plaza de Ciencias, 3\\E-28040
Madrid, Spain}

\ead{rutwig@mat.ucm.es}

\begin{abstract}

It is shown that for inhomogeneous Lie algebras having only one
Casimir operator, the latter can be explicitly constructed from
the Maurer-Cartan equations by means of wedge products. It is
further proved that this constraint imposes sharp bounds for the
dimension of the representation $R$ defining the semidirect
product. The procedure is generalized to compute also the rational
invariant of some Lie algebras.

\end{abstract}



\maketitle

\section{Introduction}

The study of (generalized) Casimir operators of symmetry groups
has acquired an importance fundamental for the understanding of
physical theories, where they constitute a valuable tool for
classification schemes and the subsequent establishment of mass
formulae. The invariants of a Lie algebra serve principally to
label irreducible representations, and are therefore of interest
to separate multiplets, their eigenvalues being identified with
the quantum numbers of certain physical observables. Among the
different physical situations where Lie algebras appear, like in
geometric symmetry groups related to degeneracies of energy levels
or spectrum generating algebras associated to non-invariance
groups, the most important case probably corresponds to those
symmetries depending on the potential energy functions (e.g.
Coulomb problem), nowadays known as dynamical symmetry groups. For
these the invariants are specially useful, since they enable us to
express the Hamiltonian of the system in terms of the Casimir
operators of a Lie algebra and some of its Lie subalgebras (e.g.,
the Interaction Boson Model) \cite{Ia,He,Nee}. Other situations
involving the invariants of an algebra and certain of its
subalgebras appear in connection with symmetry breaking problems
\cite{Boh}.

\medskip

The question how to find the invariants of a Lie algebra has been
approached by many different methods, from the study of enveloping
algebras of semisimple Lie algebras to analytical methods based on
the integration of a set of linear first order partial
differential equations \cite{Be,Ro,Per,AA,Zas,De2,Pe,Az1,C23}.
Other methods, like tensor operator techniques or algebraic
reductions, have been successfully developed for special classes
of Lie algebras \cite{C33,Bo,WW}. Another possibility to obtain
the Casimir operators, or more generally, the invariants by the
coadjoint representation, is the use of the structural or
Maurer-Cartan equations of the Lie algebra. Differential forms,
being a powerful tool for many physical problems, have however not
been used extensively in the invariant problem, although they are
a common technique in the study of Poisson structures and related
integrability conditions \cite{He,Ber,Yud}. This interpretation is
very close to that of total differential equations \cite{Pe}, and
suggests that for Lie algebras with one Casimir operator, the
latter could be somehow encoded in the Maurer-Cartan equations, by
means of wedge products. On the other hand, algebras with only one
invariant cannot be obtained by contraction of semisimple Lie
algebras (up to the trivial case in dimension three \cite{WW}),
thus the invariant has to be computed directly.

\medskip

In this work we prove that for semidirect products of semisimple
and Abelian Lie algebras having only one invariant for the
coadjoint representation, the Casimir operator of the algebra is
completely determined by the corresponding Maurer-Cartan
equations. As a consequence, such algebras will be endowed with a
concrete geometrical structure. We also point out that the
procedure enables us to find the commuting polynomials for the
cases where the Lie algebras has a rational invariant, and even
allows to recover the invariants for Lie algebras having two
Casimir operators. These results constitute a complement to the
geometrical method developed recently in \cite{Bo,Bo2} for Lie
algebras with certain structure. As applications, we derive a
basis independent expression for the Casimir operator of the
special affine Lie algebras $\frak{sa}(n\mathbb{R})$ used in
affine quantum gauge theories \cite{He}.

\medskip

Any Lie algebra $\frak{g}$ and any representation $R$ considered
in this work  is defined over the field $\mathbb{R}$ of real
numbers. We convene that non-written brackets are either zero or
obtained by antisymmetry. We also assume implicitly the Einstein
summation convention. Abelian Lie algebras of dimension $n$ will
be denoted by $nL_{1}$.

\section{Invariants of Lie algebras. Maurer-Cartan equations}

Among the  multiple algebraic and analytical methods developed in
the literature in order to determine the Casimir invariants of Lie
algebras, the procedure based on partial differential equations
(PDEs) has probably become the most standard \cite{Zas,Pe}. Given
a basis  $\left\{X_{1},..,X_{n}\right\}  $   of $\frak{g}$
and the structure tensor $\left\{  C_{ij}%
^{k}\right\}  $, we consider the realization of $\frak{g}$ in the
space $C^{\infty}\left(  \frak{g}^{\ast }\right) $ determined by
the differential operators:
\begin{equation}
\widehat{X}_{i}=C_{ij}^{k}x_{k}\frac{\partial}{\partial x_{j}},
\end{equation}
where $\left[  X_{i},X_{j}\right]  =C_{ij}^{k}X_{k}$ $\left( 1\leq
i<j\leq n,\;1\leq k\leq n\right)  $. It is straightforward to
verify that  the brackets $\left[ \widehat
{X}_{i},\widehat{X}_{j}\right] =-C_{ij}^{k}\widehat{X}_{k}$ are
satisfied, showing that they define a linear representation of the
algebra. Casimir operators of $\frak{g}$ correspond to functions
on the generators $F\left( X_{1},..,X_{n}\right)  $ of $\frak{g}$
that satisfy the constraint
\begin{equation}
\left[  X_{i},F\left(  X_{1},..,X_{n}\right)  \right]=0.
\end{equation}
Using the ansatz by PDEs, it can be seen that Casimir operators
constitute a special case corresponding to polynomial solutions to
the system of linear first order partial differential
equations:%
\begin{equation}
\widehat{X}_{i}F\left(  x_{1},..,x_{n}\right)  =C_{ij}^{k}x_{k}%
\frac{\partial F}{\partial x_{j}}\left(  x_{1},..,x_{n}\right)
=0,\;1\leq i\leq n.\label{sys}
\end{equation}
Solutions of (\ref{sys}) are called generalized Casimir invariants
or invariants of the coadjoint representation of $\frak{g}$. The
classical Casimir operators are recovered using the symmetrization
map $Sym(x_{1}^{a_{1}}..x_{p}^{a_{p}})=\frac{1}{k!}
\sum_{\sigma\in S_{p}}
x_{\sigma(1)}^{a_{1}}...x_{\sigma(p)}^{a_{p}}$ and then replacing
the variables $x_{i}$ by the corresponding generator $X_{i}$
\cite{AA}. A maximal set of functionally independent solutions of
$\left( {3}\right) $ will be called a fundamental set of
invariants. The number $\mathcal{N}(\frak{g})$ of independent
solutions to the system is given by \cite{Be}:
\begin{equation}
\mathcal{N}(\frak{g})=\dim \,\frak{g}-{\rm sup}_{x_{1},..,x_{n}}
{\rm rank} \left( C_{ij}^{k}x_{k}\right),\label{BBB}
\end{equation}
where $A(\frak{g}):=\left(C_{ij}^{k}x_{k}\right)$ is  the matrix
which represents the commutator table over the basis
$\left\{X_{1},..,X_{n}\right\}$. Some years ago, a method based on
total differential equations was proposed in \cite{Pe}. For
algebras with one invariant, system (\ref{sys}) is simplified to a
total differential equation of the type:
\begin{equation}
dF=dx_{1}+U_{12}dx_{2}+...+U_{1n}dx_{n}=0,\label{RP1}
\end{equation}
by means of successive reductions, where the $U_{1i}$ are
functions of the generators of $\frak{g}$ obtained from the
commutator matrix $A(\frak{g})$ \cite{Pe}. The solution of
(\ref{RP1}) is thus a first integral $F=\sum_{i=1}^{n}f_{i}x_{i}$
of equation (\ref{RP1}), the $f_{i}$ being the result of deleting
the common factors in the $U_{i}$. This method reduced the
computation of the Casimir operator to the evaluation of
$\dim(\frak{g})-1$ determinants and the integration of the total
differential equation (\ref{RP1}). This ansatz can be easily
reformulated in terms of differential forms, which constitute a
more natural frame for this type of algebras. In terms of the
Maurer-Cartan equations, the Lie algebra $\frak{g}$
is described as follows: Given the structure tensor $\left\{  C_{ij}%
^{k}\right\}  $ over the basis $\left\{  X_{1},..,X_{n}\right\} $,
the identification of the dual space $\frak{g}^{\ast}$ with the
left-invariant Pfaffian forms on the simply connected Lie group
whose algebra is isomorphic to $\frak{g}$ allows to define an
exterior differential $d$ on $\frak{g}^{\ast}$ by
\begin{equation}
d\omega\left(  X_{i},X_{j}\right)  =-C_{ij}^{k}\omega\left(  X_{k}\right)
,\;\omega\in\frak{g}^{\ast}.
\end{equation}
With this operator coboundary $d$ we can rewrite $\frak{g}$ as a closed system of $2$-forms%
\begin{equation}
d\omega_{k}=-C_{ij}^{k}\omega_{i}\wedge\omega_{j},\;1\leq i<j\leq\dim\left(
\frak{g}\right)  ,
\end{equation}
called the Maurer-Cartan equations of $\frak{g}$. In particular,
the condition $d^{2}\omega_{i}=0$ for all $i$ is equivalent to the
Jacobi condition. In order to relate this approach with the number
of invariants, we consider the linear subspace
$\mathcal{L}(\frak{g})=\mathbb{R}\left\{ d\omega_{i}\right\}
_{1\leq i\leq \dim\frak{g}}$ of $\bigwedge^{2}\frak{g}^{\ast}$
generated by the $2$-forms $d\omega_{i}$ \cite{Ca43}. It follows
at once that $\dim\mathcal{L}(\frak{g})=\dim\left( \frak{g}\right)
$ if and only if $d\omega _{i}\neq0$ for all $i$, that is, if the
condition $\dim\left( \frak{g}\right)  =\dim\left[
\frak{g},\frak{g}\right] $ holds. If
$\omega=a^{i}d\omega_{i}\,\;\left( a^{i}\in\mathbb{R}\right)  $ is
a generic element of $\mathcal{L}(\frak{g})$, then we can find
$j_{0}\left( \omega\right)  \in\mathbb{N}$ such that
\begin{equation}
\bigwedge^{j_{0}\left(  \omega\right)  }\omega\neq0,\quad \bigwedge
^{j_{0}\left(  \omega\right)  +1}\omega\equiv0.
\end{equation}
The quantity $j_{0}\left(  \frak{g}\right)  $ defined by
\begin{equation}
j_{0}\left(  \frak{g}\right)  =\max\left\{  j_{0}\left(
\omega\right) \;|\;\omega\in\mathcal{L}(\frak{g})\right\}.
\label{MCa1}
\end{equation}
constitutes a numerical invariant of the Lie algebra $\frak{g}$,
and allows to rewrite equation (\ref{BBB}) as
\begin{equation}
\mathcal{N}\left(  \frak{g}\right)  =\dim\frak{g}-2j_{0}\left(  \frak{g}%
\right). \label{BB1}
\end{equation}
This identity implies that $j_{0}\left(  \frak{g}\right)$
coincides with the number of internal  labels necessary to
describe a general irreducible representation of $\frak{g}$
\cite{Ca43,Ra,Sh}.

\smallskip

This reformulation in terms of differential forms can be useful to
construct the Casimir operator when the condition
$\mathcal{N}(\frak{g})=1$ is satisfied. As an example how it can
be computed from the Maurer-Cartan
equations, consider the 5-dimensional Lie algebra $\frak{g}=\frak{sl}%
\left(  2,\mathbb{R}\right)
\overrightarrow{\oplus}_{D_{\frac{1}{2}}}2L_{1}$
given by the brackets%
\begin{equation*}%
\begin{array}
[c]{llll}%
\left[  X_{1},X_{2}\right]  =2X_{2}, & \left[  X_{1},X_{3}\right]
=-2X_{3} &
\left[  X_{2},X_{3}\right]  =X_{1} & \left[  X_{1},X_{4}\right]  =X_{4},\\
\;\left[  X_{1},X_{5}\right]  =-X_{5}, & \left[
X_{2},X_{5}\right]  =X_{4}, & \left[  X_{3},X_{4}\right]  =X_{5}.
&
\end{array}
\end{equation*}
The structure equations are easily seen to be%
\begin{equation}
\fl \begin{array}[c]{lll} d\omega_{1}=-\omega_{2}\wedge\omega_{3},
& d\omega_{2}=-2\omega_{1}\wedge
\omega_{2}, & d\omega_{3}=2\omega_{1}\wedge\omega_{3},\\
d\omega_{4}=-\omega_{1}\wedge\omega_{4}-\omega_{2}\wedge\omega_{5},
&
d\omega_{5}=\omega_{1}\wedge\omega_{5}-\omega_{3}\wedge\omega_{4}.
&
\end{array}
\end{equation}
Let us consider a generic element $\omega\in\frak{g}^{\ast}$:
$\omega=a^{1}\omega_{1}+a^{2}\omega_{2}+a^{3}\omega_{3}+a^{4}\omega_{4}
+a^{5}\omega_{5}$, where $a^{i}\in\mathbb{R}$ are considered as
variables. The coboundary operator $d$ is given by:
\begin{eqnarray*}
d\omega
=&a^{1}\omega_{2}\wedge\omega_{3}+2a^{2}\omega_{1}\wedge\omega
_{2}-2a^{3}\omega_{1}\wedge\omega_{3}+a^{4}\left(
\omega_{1}\wedge\omega
_{4}+\omega_{2}\wedge\omega_{5}\right)  +\\
 &a^{5}\left(-\omega_{1}\wedge\omega_{5}+\omega_{3}\wedge\omega_{4}\right) .
\end{eqnarray*}
Now, computing the wedge product $\omega\wedge d\omega\wedge
d\omega$ and expanding it, we get the expression
\begin{equation}
\fl \omega\wedge d\omega\wedge d\omega=-6\left(  a^{1}a^{4}a^{5}+a^{2}(a^{5}%
)^{2}-a^{3}(a^{4})^{2}\right)
\omega_{1}\wedge\omega_{2}\wedge\omega
_{3}\wedge\omega_{4}\wedge\omega_{5}.
\end{equation}
Observe that the polynomial $\Phi=a^{1}a^{4}a^{5}+a^{2}(a^{5})^{2}-a^{3}%
(a^{4})^{2}$ is homogeneous of degree two in the variables $a^{4}$ and $a^{5}%
$, corresponding to the generators in the radical of $\frak{g}$.
Now, replacing the variables $a^{i}$ by
the corresponding $x_{i}$, it is straightforward to verify that $C=x_{1}%
x_{4}x_{5}+x_{2}x_{5}^{2}-x_{3}x_{4}^{2}$ satisfies the system
(\ref{sys}) corresponding to this Lie algebra. Therefore the
symmetrized polynomial is the Casimir operator of $\frak{g}$.

\medskip

The main objective of this work is to show the correctness of this
observation for algebras satisfying the constraint
$\frak{g}=[\frak{g},\frak{g}]$, i.e., being perfect. This apparent
restriction is necessary, since for perfect Lie algebras the
existence of complete sets of invariants formed by polynomials is
ensured \cite{AA}, while for non-perfect algebras we can even have
transcendental invariants \cite{AA,Bo2,Sh}. However, we will show
that the Maurer-Cartan are also useful to find the commuting
polynomials that constitute the rational invariant for Lie
algebras with a codimension one commutator subalgebra. This also
allows a method to find the Casimir operators of various Lie
algebras with two invariants.

\section{Inhomogeneous Lie algebras}

Among all Lie algebras having a non-trivial Levi decomposition,
inhomogeneous Lie algebras, i.e., semidirect products
$\frak{g}=\frak{s}\overrightarrow{\oplus}_{R}(\dim R)L_{1}$ of
semisimple and Abelian Lie algebras, have an important property
concerning the structure of their Casimir invariants. As already
told, it is known that these algebras admit polynomial invariants
whenever  the constraint $\left[\frak{g},\frak{g}\right]=\frak{g}$
is satisfied \cite{AA}. It is natural to ask whether the Levi
decomposition induces some kind of homogeneity of the Casimir
operators with respect to the variables of the Levi subalgebra
$\frak{s}$ and the variables of the Abelian radical. This
property, which holds exclusively for this class, is a consequence
of the so-called missing label problem associated to these
algebras \cite{Sh} and the natural contraction related to it
\cite{C72,C73}:

\begin{theorem} Let
$\frak{g}=\frak{s}\overrightarrow{\oplus}_{R}(\dim R )L_{1}$. Then
any Casimir invariant is an homogeneous polynomial in the
variables of the radical $(\dim R)L_{1}$.
\end{theorem}

We also remark that the particular structure of the invariants of
inhomogeneous Lie algebras is deeply connected with the expansion
method considered by Rosen in \cite{Ro} for unitary and
pseudo-orthogonal algebras. Actually the previous homogeneity with
respect to the appropriate variables is one of the principal
reasons for the possibility of recovering the Casimir operators of
semisimple Lie algebras from those of an inhomogeneous
contraction.

\medskip

We now focus on  inhomogeneous Lie algebras satisfying the
constraint $\mathcal{N}(\frak{g})=1$. For this case, various
interesting structural properties emerge, that will allow us to
find an intrinsic construction of the Casimir operator.

\begin{lemma}
Let $\mathcal{L=}\left\{  \omega_{1},..,\omega_{2r+1}\right\}  $
be a system
of independent 1-forms such that $d\omega_{i}\in\bigwedge^{2}\mathcal{L}%
-\left\{  0\right\}  $ for $i=1..2r+1$. Let $\omega=\sum_{i=1}^{2r+1}%
a_{i}\omega_{i}\in\mathcal{L}$ be an element such that
\[
\left(  \bigwedge^{r}d\omega\right)  \wedge\omega=\Phi\left(  \,a_{1}%
,..,a_{2r+1}\right)
\omega_{1}\wedge\omega_{2}\wedge...\wedge\omega _{2r+1}\neq0,
\]
where $\Phi\left(  \,a_{1}%
,..,a_{2r+1}\right)$ is a polynomial in the variables $a_{i}$.
Then
\begin{equation}
\bigwedge^{r}d\omega=\frac{1}{r+1}\sum_{i=1}^{2r+1}\left(
-1\right) ^{i+1}\frac{\partial\Phi}{\partial
a_{i}}\,\omega_{1}\wedge...\wedge
\widehat{\omega_{i}}\wedge...\wedge\omega_{2r+1},\label{EAP}
\end{equation}
where $\widehat{\omega_{i}}$ denotes omission of the element
$\omega_{i}$.
\end{lemma}

\begin{proof}
From the properties of the wedge product it follows at once that
$\Phi\left( \,a_{1},..,a_{2r+1}\right)  $ is a homogeneous
polynomial of degree $r+1$ in the variables $a_{1},..,a_{2r+1}$.
Define the $2r$-form \begin{equation}
\theta=\sum_{i=1}^{2r+1}\left( -1\right)
^{i+1}\frac{\partial\Phi}{\partial
a_{i}}\,\omega_{1}\wedge...\wedge\widehat{\omega_{i}}\wedge...\wedge
\omega_{2r+1}.
\end{equation}
By assumption, $\Phi\neq 0$, thus the 2-form $\theta$ is
non-vanishing. Considering the 1-form
$\omega=\sum_{i=1}^{2r+1}a_{i}\omega_{i}$ and taking the wedge
product of $\theta$ and $\omega$, we obtain that
\begin{eqnarray}
\theta\wedge\omega & =\sum_{i=1}^{2r+1}\left(  -1\right)
^{i+1}\frac
{\partial\Phi}{\partial a_{i}}\,\omega_{1}\wedge...\wedge\widehat{\omega_{i}%
}\wedge...\wedge\omega_{2r+1}\wedge\left(
\sum_{i=1}^{2r+1}a_{i}\omega
_{i}\right)  \nonumber\\
& =\sum_{i=1}^{2r+1}\left(  -1\right)
^{i+1}a_{i}\frac{\partial\Phi}{\partial
a_{i}}\,\omega_{1}\wedge...\wedge\widehat{\omega_{i}}\wedge...\wedge
\omega_{2r+1}\wedge\omega_{i}\nonumber\\
& =\sum_{i=1}^{2r+1}\left(  -1\right)
^{2r+2}a_{i}\frac{\partial\Phi }{\partial
a_{i}}\,\omega_{1}\wedge...\wedge\omega_{i}\wedge...\wedge
\omega_{2r+1}.\label{KP1}
\end{eqnarray}
By homogeneity of the polynomial $\Phi$, the Euler identities
imply the equality
\begin{equation}
\sum_{i=1}^{2r+1}a_{i}\frac{\partial\Phi}{\partial a_{i}}=\left(
r+1\right) \,\Phi,
\end{equation}
from which the assertion follows, since
$\left((r+1)\bigwedge^{r}d\omega-\theta\right)\wedge \omega=0$ by
equation (\ref{KP1}).
\end{proof}

\medskip

This result is formally very close to the approach by means of
total differential equations as presented in (\ref{RP1}). We will
later see that actually the use of wedge products determines the
Casimir operator, corresponding to the first integrals of
(\ref{RP1}). Moreover, in contrast to the general case, the
constraint of having only one Casimir invariant imposes some
restrictions on the possible dimension of a representation. The
following results establish (sharp) bounds for the dimension of
the radical in inhomogeneous Lie algebras with only one invariant.

\begin{lemma}
Let $\frak{g}=\frak{s}\overrightarrow{\oplus}_{R}nL_{1}$ be an
indecomposable Lie algebra. If $\mathcal{N}\left(  \frak{g}\right)
=1$, then $\dim R=n\geq \mathrm{rank}\;\frak{s}+1$.
\end{lemma}

\begin{proof}
We consider the Levi subalgebra $\frak{s}$ of $\frak{g}$. It is
known that, in addition to the Casimir operators of these
algebras, we need
\begin{equation}
\frac{1}{2}\left(
\dim\frak{g}-\mathcal{N}(\frak{g})-\dim\frak{h}-\mathcal{N}(\frak{h})\right)+l^{\prime}
\label{ML}
\end{equation}
additional operators (commonly called missing label operators) in
order to label unambigously the states of $\frak{g}$ with respect
to the subalgebra $\frak{s}$ \cite{Sh}. For this reduction,
$l^{\prime}$ is clearly zero, thus we need
\begin{equation}
\frac{1}{2}\left(\dim\frak{s}+n-1-\dim\frak{s}-\mathrm{rank}\;\frak{s}\right)=
\frac{1}{2}\left(n-1-\mathrm{rank}\;\frak{s}\right)\geq 0,
\end{equation}
additional operators. The latter being a non-negative quantity, we conclude
that $\dim R=n\geq 1+\mathrm{rank}\;\frak{s}$.
\end{proof}

\smallskip

We observe that the example presented at the beginning exactly
satisfies the equality, showing that the lower bound is sharp. As
to upper bounds, the result is a consequence of the specific
structure of system (\ref{sys}) for these algebras:

\begin{lemma}
Let $\frak{g}=\frak{s}\overrightarrow{\oplus}_{R}nL_{1}$ be an
indecomposable Lie algebra. If $\mathcal{N}\left(  \frak{g}\right)
=1$, then $\dim R=n\leq\dim\frak{s}+1$.
\end{lemma}

\begin{proof}
Consider a basis $\left\{ X_{1},..,X_{r},Y_{1},..,Y_{n}\right\} $
 of $\frak{g}$ such that $\left\{ X_{1},..,X_{r}\right\}
$ is a basis of the Levi part $\frak{s}$ and $\left\{
Y_{1},..,Y_{n}\right\}  $ is a basis of the representation space
$R$.\footnote{Since the radical is Abelian, we can naturally
identify it with the representation space $R$, where $n=\dim R$.}
 The invariant of $\frak{g}$ is obtained from the
system (\ref{sys}), expressed   in matrix form:%
\begin{equation}
\left(
\begin{array}
[c]{c}%
\widehat{X}_{i}F\\
\widehat{Y}_{i}F
\end{array} \right)=
\left(
\begin{array}
[c]{cc}%
C_{ij}^{k}x_{k} & C_{ij}^{k}y_{k}\\
-C_{ij}^{k}y_{k} & 0
\end{array}
\right)  \left(
\begin{array}
[c]{c}%
\frac{\partial F}{\partial x_{i}}\\
\frac{\partial F}{\partial y_{i}}%
\end{array}
\right)  =0.
\end{equation}
In particular, the number $\mathcal{M}\left(  \frak{g}\right)  $
of invariants which depend only on the variables $\left\{
y_{1},..,y_{n}\right\}  $ is given by:
\begin{equation}
\mathcal{M}\left(  \frak{g}\right)  =\dim R-\mathrm{rank\,}\left(  C_{ij}%
^{k}y_{k}\right)  ,\label{SPEV}
\end{equation}
where $B=\left(  C_{ij}^{k}y_{k}\right)  $ is a $\left( n\times
r\right)  $-matrix describing the action of the Levi part
$\frak{s}$ on the radical $R$ \cite{C23}. In any case, the following
inequality holds:
\begin{equation}
\mathrm{rank\,}\left(  C_{ij}^{k}y_{k}\right)  \leq\min\left\{
n,r\right\}  .\label{RA}
\end{equation}
In view of this bound, there are only two possibilities:

\begin{enumerate}
\item  If $\mathcal{M}\left(  \frak{g}\right)  =0$, then $\dim
R=n=\mathrm{rank\,}\left(  C_{ij}^{k}y_{k}\right)  $, which
implies that $\dim R\leq\dim\frak{s}$, as the condition $\dim
R>\dim\frak{g}$ would contradict inequality (\ref{RA}). Further,
since equality of the dimensions of $R$ and $\frak{s}$ is
forbidden by parity, we conclude that
\begin{equation}
\dim R\leq\dim\frak{s}-1.
\end{equation}

\item  If $\mathcal{M}\left(  \frak{g}\right)  =1$, then the
Casimir operator of $\frak{g}$ depends only on variables of the
Abelian radical. In this case
\begin{equation} 1=\dim
R-\mathrm{rank\,}\left( C_{ij}^{k}y_{k}\right)  ,
\end{equation}
and the only relevant case $n>r$ implies that $\mathrm{rank}\left(
C_{ij}^{k}y_{k}\right)=n-1\leq r=\dim\frak{s}<n$ and thus that
\begin{equation}
\dim R\leq\dim\frak{s}+1.
\end{equation}
\end{enumerate}
\end{proof}

As a consequence of these lemmas, the representation $R$
describing the semidirect product
$\frak{g}=\frak{s}\overrightarrow{\oplus}_{R}nL_{1}$ with
$\mathcal{N}(\frak{g})=1$ satisfies ${\rm rank } \frak{s}+1\leq
\dim R\leq \dim\frak{s}-1$ or $\dim R =\dim\frak{s}+1$. The latter
constitutes an isolated type of algebras, and it follows from the
proof that their Casimir operator depends only on the variables of
the radical \cite{C23}. The other possibility, $\dim R<\dim
\frak{s}$, presents additional features that are
 of relevance for the Casimir operator. Like before,
suppose that  $\left\{  X_{1},..,X_{r}%
,Y_{1},..,Y_{n}\right\}  $ is a basis of $\frak{g}$ such that
$\left\{ X_{1},..,X_{r}\right\}  $ is a basis of   $\frak{s}$ and
$\left\{  Y_{1},..,Y_{n}\right\}  $ a basis of the radical
$nL_{1}$. Denote the dual basis by $\left\{
\omega_{1},..,\omega_{r},\theta_{1},..,\theta _{n}\right\}  $.

\begin{lemma}
If there exists an element
$\omega\in\frak{g}^{\ast}$ such that
\begin{equation}
\fl \left(  \bigwedge^{j_{0}\left(  \frak{g}\right)
}d\omega\right) \wedge
\omega=\Phi\left(  \,a_{1},..,a_{r},b_{1},..b_{n}\right)  \omega_{1}%
\wedge...\wedge\omega_{r}\wedge\theta_{1}\wedge...\wedge\theta_{n}\neq0,\label{Kar1}
\end{equation}
then $\Phi\left(  a_{i},b_{j}\right)  $ is a homogeneous
polynomial in both the variables $\left\{  a_{i}\right\}  $ and
$\left\{  b_{j}\right\}  $.
\end{lemma}

\begin{proof}
Since $\mathcal{N}\left(  \frak{g}\right)  =1$, we obtain from
(\ref{BB1}) that
\begin{equation}
j_{0}\left(  \frak{g}\right)  =\frac{1}{2}\left(  n+r-1\right)
.\label{JLA}
\end{equation}
For the Levi subalgebra $\frak{s}$ we have the equality
$\dim\frak{s}=r={\rm rank }\frak{s}+2j_{0}\left( \frak{s}\right)
$, and insertion into equation (\ref{JLA}) gives
\begin{equation}
j_{0}\left(  g\right)  =j_{0}\left(  \frak{s}\right)
+\frac{1}{2}\left( {\rm rank }\frak{s}-1+n\right)  .
\end{equation}
Let us consider a generic element $\omega\in\frak{g}^{\ast}$:
\begin{equation}
\omega=\sum_{i=1}^{r}a_{i}\omega_{i}+\sum_{j=1}^{n}b_{j}\theta_{j}%
,
\end{equation}
where $\;a_{i},b_{j}\in\mathbb{R}$ are arbitrary constants. The exterior differential $d$ is
determined by:
\begin{equation}
d\omega=\sum_{i=1}^{r}a_{i}d\omega_{i}+\sum_{j=1}^{n}b_{j}d\theta_{j}.
\end{equation}
Wedge products of the latter $2$-form can be rewritten as
\begin{equation}
\bigwedge^{p}d\omega=\bigwedge^{p}\left(  \xi_{1}+\xi_{2}\right)
=\sum _{k=1}^{p}\left(
\begin{array}
[c]{c}
p \\
k
\end{array}\right)
\bigwedge^{k}\xi_{1}\wedge\bigwedge^{p-k}\xi_{2},\label{JLB}
\end{equation}
where $\xi_{1}=$ $\sum_{i=1}^{r}a_{i}d\omega_{i}$ and $\xi_{2}=\sum_{j=1}%
^{n}b_{j}d\theta_{j}$. In particular, any summand of (\ref{JLB})
is a wedge product of $k+p$ elements $\omega_{i}$ and $p-k$
elements $\theta_{j}$. We say that $\bigwedge^{p}\left(
\xi_{1}+\xi_{2}\right)  $ has bi-degree $\left( k+p,p-k\right)  $
in the $\left(  \frak{s},R\right)  $-variables. It follows at once
from this that any term of $\bigwedge^{p}\left(  \xi_{1}+\xi
_{2}\right)  $ has degree $k$ in the variables $a_{i}$, and degree
$p-k$ in the $b_{i}$'s. Further, it is clear from (\ref{BB1}) that
$\bigwedge^{p}\xi_{1}=0$ for any $p>j_{0}\left(  \frak{s}\right)$,
thus
\begin{equation}
\bigwedge^{j_{0}\left(  \frak{g}\right)  }\left(
\xi_{1}+\xi_{2}\right) =\sum_{k\geq0}\left(
\begin{array}
[c]{c}%
j_{0}\left(  \frak{g}\right)  \\
k
\end{array}\right)\bigwedge^{j_{0}\left(
\frak{s}\right)  -k}\xi_{1}\wedge\bigwedge^{\frac{1}{2}\left(  \mathrm{rank}%
\frak{s}-1+n\right)  +k}\xi_{2}.\label{JLC}
\end{equation}
By assumption, $\bigwedge^{j_{0}\left(  \frak{g}\right)
}d\omega\wedge \omega\neq0$, thus applying the preceding lemma
we obtain the decomposition:
\begin{eqnarray}
\fl \bigwedge^{j_{0}\left(  \frak{g}\right)  }\left(
\xi_{1}+\xi_{2}\right)    =\frac{1}{j_{0}\left(  \frak{g}\right)
+1}\sum_{i=1}^{r}\left(  -1\right)
^{i+1}\frac{\partial\Phi}{\partial
a_{i}}\,\omega_{1}\wedge...\wedge
\widehat{\omega_{i}}\wedge...\wedge\omega_{r}\wedge\theta_{1}\wedge
...\wedge\theta_{i}\wedge...\wedge\theta_{n}\nonumber\\
\lo   +\frac{1}{j_{0}\left(  \frak{g}\right)
+1}\sum_{i=1}^{n}\left( -1\right)
^{n+i+1}\frac{\partial\Phi}{\partial
b_{i}}\,\omega_{1}\wedge...\wedge
\omega_{i}\wedge...\wedge\omega_{r}\wedge\theta_{1}\wedge...\wedge
\widehat{\theta_{i}}\wedge...\wedge\theta_{n}.\nonumber\\
  \label{JLD}
\end{eqnarray}
The bi-degree of the terms in the latter product are either
$\left( r-1,n\right)  $ or $\left(  r,n-1\right)  $, since the
bi-degree of $\Phi$ is $\left(  r,n\right)  $. In view of equation
(\ref{EAP}), this decomposition is possible only if
\begin{equation}
\fl \bigwedge^{j_{0}\left(  \frak{g}\right)  }\left(
\xi_{1}+\xi_{2}\right) =\left(
\begin{array}
[c]{c}%
j_{0}\left(  \frak{g}\right)  \\
\frac{n-\mathrm{rank}\frak{s}-1}{2}
\end{array}\right)\bigwedge^{\frac{\dim\frak{s}-n-1}{2}}\xi_{1}\wedge\bigwedge^{n}\xi
_{2}+\left(\begin{array}
[c]{c}%
j_{0}\left(  \frak{g}\right)  \\
\frac{n-\mathrm{rank}\frak{s}-1}{2}
\end{array}\right)\bigwedge^{\frac{\dim\frak{s}-n+1}{2}}\xi_{1}\wedge\bigwedge^{n-1}%
\xi_{2},\label{JLE}
\end{equation}
because these are the only products having the required
bi-degrees. Finally, considering the wedge product of (\ref{JLE})
with $\omega$, the following identities can be easily verified:
\begin{eqnarray}
\fl \bigwedge^{\frac{\dim\frak{s}-n-1}{2}}\xi_{1}\wedge\bigwedge^{n}\xi_{2}%
\wedge\left(
\sum_{i=1}^{r}a_{i}\omega_{i}+\sum_{j=1}^{n}b_{j}\theta
_{j}\right)    &
=\bigwedge^{\frac{\dim\frak{s}-n-1}{2}}\xi_{1}\wedge
\bigwedge^{n}\xi_{2}\wedge\left(  \sum_{i=1}^{r}a_{i}\omega_{i}\right)  ,\label{JLF1}\\
\fl
\bigwedge^{\frac{\dim\frak{s}-n+1}{2}}\xi_{1}\wedge\bigwedge^{n-1}\xi
_{2}\wedge\left(
\sum_{i=1}^{r}a_{i}\omega_{i}+\sum_{j=1}^{n}b_{j}\theta
_{j}\right)    &
=\bigwedge^{\frac{\dim\frak{s}-n+1}{2}}\xi_{1}\wedge
\bigwedge^{n-1}\xi_{2}\wedge\left(
\sum_{j=1}^{n}b_{j}\theta_{j}\right)  .\label{JLF2}
\end{eqnarray}
This shows that any term of  $\Phi$ has degree $n$ in the
variables $\left\{ b_{j}\right\}  $ and degree $\frac{r-n+1}{2}$
in the variables $a_{i}$, and therefore that the polynomial $\Phi$
is homogeneous in these variables.
\end{proof}

This result is a sharpened version of Theorem 1, and constitutes the essential
step to recover the Casimir operator intrinsically.

\section{Construction of the Casimir operator from the Maurer-Cartan equations}

In this section we prove that for inhomogeneous Lie algebras
possessing only one Casimir operator, the latter can be
constructed using only the Maurer-Cartan equations, basing on the
homogeneity properties previously seen. In \cite{C33} a result of
similar nature was proposed, using a special extension of degree
one of the algebra. We will use this result for a stronger
statement, for which reason we briefly recall the result:

\begin{proposition}
Let $\frak{s}\overrightarrow{\oplus}_{R}nL_{1}$ a perfect Lie
algebra such that
$\mathcal{N}(\frak{s}\overrightarrow{\oplus}_{R}nL_{1})=1$. Let
$C$ be the invariant of minimal degree. Then
$\frak{s}\overrightarrow{\oplus}_{R}nL_{1}$ admits an extension
$\widehat{\frak{g}}$ of degree one satisfying
$\mathcal{N}(\widehat{\frak{g}})=0$ and such that
$\left|A(\frak{g})\right|= C^{2}$ is the square power of $C$.
\end{proposition}

Expressed in terms of the Maurer-Cartan equations
$\left\{d\varphi_{1},...,d\varphi_{n+r+1}\right\}$ of the
extension $\widehat{\frak{g}}$, this  result establishes the existence of an element
$z^{i}d\varphi_{i}\in\mathcal{L}(\widehat{\frak{g}})$ such that
\begin{equation}
\bigwedge^{\frac{n+r+1}{2}}d\omega=\left(\frac{n+r+1}{2}\right)!\;
 C(z_{i}) \varphi_{1}\wedge...\wedge\varphi_{n+r+1}.
\end{equation}

The reason for this remarkable relation between certain extensions
and the Casimir operator of $\frak{g}$  remained however
unexplained in \cite{C33}. We will see that the Casimir operator
of $\frak{g}$, being completely determined by the Maurer-Cartan
equations themselves, implies the existence of such an extension.

\begin{theorem}
Let $\frak{g}=\frak{s}\overrightarrow{\oplus}_{R}nL_{1}$ be an
indecomposable Lie algebra with $\mathcal{N}\left( \frak{g}\right)
=1$. Then the Casimir operator $C$ of minimal degree is
intrinsically determined by the Maurer-Cartan equations of
$\frak{g}$.
\end{theorem}

\begin{proof}
We consider a basis $\mathcal{B}=\left\{
X_{1},..,X_{r},Y_{1},..,Y_{n}\right\} $, where  $\left\{
X_{1},..,X_{r}\right\}  $ is a basis of the Levi part $\frak{s}$
and $\left\{ Y_{1},..,Y_{n}\right\}  $ is a basis of the
representation space $R$. The structure tensor over $\mathcal{B}$
is given by
\begin{equation*}
\left[  X_{i},X_{j}\right]  =C_{ij}^{k}X_{k},\;\left[
X_{i},Y_{j}\right] =D_{ij}^{k}Y_{k},\;\left[  Y_{i},Y_{j}\right]
=0.
\end{equation*}
Considering the dual basis $\left\{
\omega_{1},..,\omega_{r},\theta _{1},..,\theta_{n}\right\}  $ to
$\mathcal{B}$, the Maurer-Cartan equations of $\frak{g}$ are
easily seen to be:
\begin{eqnarray}
d\omega_{i}  &  =C_{jk}^{i}\omega_{j}\wedge\omega_{k},\quad i=1..r,\nonumber\\
d\theta_{j}  &  =D_{ik}^{j}\omega_{i}\wedge\theta_{k},\quad
j=1..n.
\end{eqnarray}
By asumption, $\mathcal{N}\left(  \frak{g}\right)  =1$, thus by
formula (\ref{BB1}) the condition $j_{0}\left( \frak{g}\right)
=\frac{n+r-1}{2}$ is satisfied. Let $\widehat{\frak{g}}$ denote
the extension of $\frak{g}$ containing it as a codimension one
ideal. The Maurer-Cartan equations of $\widehat{\frak{g}}$ are
given, over the basis $\left\{
\widetilde{\omega}_{1},..,\widetilde{\omega}_{r},\widetilde{\theta
}_{1},..,\widetilde{\theta}_{n},\xi\right\}  $, by
\begin{eqnarray}
d\widetilde{\omega}_{i}  &  =d\omega_{i},\quad 1\leq i\leq r\nonumber\\
d\widetilde{\theta}_{j}  &  =d\theta_{j}+\theta_{j}\wedge\xi,\quad 1\leq j\leq n\nonumber\\
d\xi &  =0.\label{Erw1}
\end{eqnarray}
Following proposition 1, $\widehat{\frak{g}}$ satisfies
$\mathcal{N}\left( \widehat{\frak{g}}\right) =0$, which implies
the condition $ j_{0}\left(\widehat{\frak{g}}\right)
=\frac{n+r+1}{2}=j_{0}\left( \frak{g}\right)  +1$. We can
therefore find an element
$\Omega=\sum_{i=1}^{r}a_{i}d\widetilde{\omega
}_{i}+\sum_{j=1}^{n}b_{j}d\widetilde{\theta}_{j}$ such that
\begin{eqnarray}
\bigwedge^{\frac{n+r+1}{2}}\Omega&=\frac{n+r-1}{2}\; C\left(
a_{i},b_{j}\right)
\widetilde{\omega}_{1}\wedge...\wedge\widetilde{\omega}_{r}\wedge
\widetilde{\theta}_{1}\wedge...\wedge\widetilde{\theta}_{n}\wedge\xi\nonumber\\
&=\frac{n+r-1}{2}\; C\left(
a_{i},b_{j}\right)
\omega_{1}\wedge...\wedge\omega_{r}\wedge
\theta_{1}\wedge...\wedge\theta_{n}\wedge\xi,\label{Erw2}
\end{eqnarray}
where $C\left(  x_{i},y_{j}\right)  $ is the Casimir invariant of
$\frak{g}$, after replacing the $a_{i}$ by $x_{i}$ and the $b_{j}$
by $y_{j}$. We decompose the $2$-form $\Omega$ as follows:

\begin{eqnarray}
\Omega & =d\left(  \sum_{i=1}^{r}a_{i}\widetilde{\omega}_{i}+\sum_{j=1}%
^{n}b_{j}\widetilde{\theta}_{j}+\xi\right)  =d\left(  \sum_{i=1}^{r}%
a_{i}\omega_{i}+\sum_{j=1}^{n}b_{j}\widetilde{\theta}_{j}\right)  \nonumber\\
  &=d\left(  \sum_{i=1}^{r}a_{i}\omega_{i}+\sum_{j=1}^{n}b_{j}\theta_{j}\right)+\left(\sum_{j=1}^{n}b_{j}%
\theta_{j}\right)  \wedge\xi=d\omega+\left(
\sum_{j=1}^{n}b_{j}\theta _{j}\right)  \wedge\xi,\label{ErwA}
\end{eqnarray}
where $\omega=\sum_{i=1}^{r}a_{i}\omega_{i}
+\sum_{j=1}^{n}b_{j}\theta_{j}$. It follows by induction on
$k\geq1$ that wedge products of $\Omega$ can be rewritten in the
following way:
\begin{equation}
\bigwedge^{k}\Omega=\bigwedge^{k}d\omega+k\,\left(  \bigwedge^{k-1}%
d\omega\right)  \wedge\left(  \sum_{j=1}^{n}b_{j}\theta_{j}\right)
\wedge\xi.
\end{equation}
In particular, for $k=j_{0}\left(  \widehat{\frak{g}}\right)  $
the product reduces to:
\begin{equation}
\bigwedge^{j_{0}\left(  \widehat{\frak{g}}\right)  }\Omega=\frac{n+r+1}%
{2}\left(  \bigwedge^{j_{0}\left(  \frak{g}\right) }d\omega\right)
\wedge\left(  \sum_{j=1}^{n}b_{j}\theta_{j}\right)
\wedge\xi.\label{ErwB}
\end{equation}
Combining the latter   with equation (\ref{Erw2}), we get the
expression:
\begin{equation}
\fl \frac{n+r+1}{2}\left(  \bigwedge^{j_{0}\left(  \frak{g}\right)  }%
d\omega\right)  \wedge\left(  \sum_{j=1}^{n}b_{j}\theta_{j}\right)
\wedge \xi=\left(\frac{n+r+1}{2}\right)!\; C\left(  a_{i},b_{j}\right)  \omega_{1}%
\wedge...\wedge\omega_{r}\wedge\theta_{1}%
\wedge...\wedge\theta_{n}\wedge\xi.\label{ErwC}
\end{equation}
Now consider the product
\begin{equation}
\left(  \bigwedge^{j_{0}\left(  \frak{g}\right)  }d\omega\right)
\wedge \omega=\Phi\left(  a_{i},b_{j}\right)
\,\omega_{1}\wedge...\wedge\omega
_{r}\wedge\theta_{1}\wedge...\wedge\theta_{n},\label{COB}
\end{equation}
where $\Phi$ is a homogeneous polynomial of degree
$\frac{n+r+1}{2}$.\footnote{At this point it is not excluded that
this expression vanishes, i.e., that $\Phi$ is a zero polynomial.}
The decomposition of Lemma 1 implies the following identity:
\begin{eqnarray}
\bigwedge^{j_{0}\left( \frak{g}\right)  }d\omega &
=\sum_{i=1}^{r}\left( -1\right) ^{i+1}\frac{\partial\Phi}{\partial
a_{i}}\omega_{1}\wedge
..\wedge\widehat{\omega}_{i}\wedge..\wedge\omega_{r}\wedge\theta_{1}%
\wedge...\wedge\theta_{n}\nonumber\\
& + \sum_{j=1}^{n}\left(  -1\right)
^{n+j+1}\frac{\partial\Phi}{\partial
b_{j}}\omega_{1}\wedge...\wedge\omega_{r}\wedge\theta_{1}\wedge..\wedge
\widehat{\theta}_{j}\wedge..\wedge\theta_{n}.\label{WP1}
\end{eqnarray}
Taking now the wedge product
\begin{equation}
\left(  \bigwedge^{j_{0}\left(  \frak{g}\right)  }d\omega\right)
\wedge\left(  \sum_{j=1}^{n}b_{j}\theta_{j}\right)  =\sum_{j=1}^{n}b_{j}%
\frac{\partial\Phi}{\partial b_{j}}\omega_{1}\wedge...\wedge\omega_{r}%
\wedge\theta_{1}\wedge..\wedge\theta_{j}\wedge..\wedge\theta_{n},\label{WP2}
\end{equation}
and comparing it with (\ref{ErwC}), we conclude the following relation
between $\Phi$ and the Casimir operator:
\begin{equation}
\frac{n+r+1}{2}\sum_{j=1}^{n}b_{j}\frac{\partial\Phi}{\partial b_{j}}%
=\left(\frac{n+r+1}{2}\right)!\; C\left(  a_{i},b_{j}\right)
.\label{WP3}
\end{equation}
Observe that if $n=r+1$, by Lemma 2 the following constraint is
satisfied:
\begin{equation}
\frac{\partial C\left(  a_{i},b_{j}\right)  }{\partial
a_{k}}=0,\;k=1..r,
\end{equation}
and by homogeneity we conclude that $\Phi$ is a multiple of the
Casimir operator of $\frak{g}$. This proves the result for the case of
maximal dimension of the representation $R$. If $n<r$, Lemma 3 implies that
$\Phi$ is homogeneous in the variables $b_{j}$ (and therefore also
in the variables $a_{i}$). Applying the Euler theorem, we get the
identity
\begin{equation}
\sum_{j=1}^{n}b_{j}\frac{\partial\Phi}{\partial b_{j}}=\left(  \deg_{b}%
\Phi\right)  \,\Phi=n\, \Phi.\label{WP4}
\end{equation}
Inserting the latter expression into equation (\ref{ErwC}) gives
the relation
\begin{equation}
n \frac{n+r+1}{2}\,\Phi=\left(\frac{n+r+1}{2}\right)!\;
C\left(a_{i},b_{j}\right),\label{WP5}
\end{equation}
which proves that $\Phi$ is a Casimir operator of $\frak{g}$ after
replacement of $a_{i}$ by $x_{i}$ and $b_{j}$ by $y_{j}$. In
particular, for any possible dimension of $R$ we obtain that
$\Phi\neq 0$, and therefore the Casimir operator of $\frak{g}$ is
uniquely determined by the wedge product (\ref{COB}) of the
Maurer-Cartan equations.
\end{proof}

This result provides us with an algorithmic procedure to compute
the Casimir operator starting from an arbitrary basis of the
algebra. Moreover, equation (\ref{COB}) shows that performing
wedge products of the defining equations of $\frak{g}$ and
extending it to an algebra satisfying the condition
$\mathcal{N}(\widehat{\frak{g}})=0$ leads to the same result. We
remark that the key point is the homogeneity of the Casimir
operator with respect to the variables of the Levi subalgebra and
the radical. Incidentally, this method implies the existence of a
supplementary geometrical property.

\medskip

Contact forms constitute, in some sense, an analogous concept to
symplectic forms for odd dimensional manifolds. Although generally
weaker than the symplectic frame of Hamiltonian Mechanics, contact
structures appear naturally in many physical problems, like
generalizations of magnetic monopoles \cite{Ito}, irreversible
thermodynamical systems \cite{Mru} or geometric formulations of
gravity coupled with Yang-Mills fields \cite{Vig}. We recall that
a linear contact form on a Lie algebra $\frak{g}$ of dimension
$2m+1$ is an element $\omega\in\frak{g}^{*}$ such that
$\omega\wedge \left(\bigwedge^{m} d\omega\right)\neq 0$. In
particular, formula (\ref{BB1}) implies that
$\mathcal{N}(\frak{g})=1$, although the converse is not
necessarily true.

\begin{proposition}
If $\frak{g}=[\frak{g},\frak{g}]=\frak{s}\overrightarrow{\oplus}_{R}nL_{1}$ is
indecomposable and satisfies $\mathcal{N}\left( \frak{g}\right)
=1$, then $\frak{g}$ is endowed with a linear contact form.
\end{proposition}

The proof follows easily from the previous argumentation. Again,
two cases must be considered according to the dimension of the
Abelian radical. If $n=r+1$, equation (\ref{WP3}) shows that the
1-form defined by $\xi=\sum_{j=1}^{n}b_{j}\theta_{j}$ is a contact
form. In this case, equations (\ref{WP1}) and (\ref{WP2}) imply
the relation
\begin{equation}
\bigwedge^{\frac{n+r-1}{2}}d\xi \wedge
\xi=\left(\frac{n+r}{2}\right)!\; C(b_{j}) \omega_{1}\wedge..
\wedge \omega_{r}\wedge\theta_{1}\wedge..\wedge \theta_{n}.
\label{KoFa}
\end{equation}
On the other hand, if $n<r$, equation (\ref{WP4}) establishes that
$\sum_{j=1}^{n}b_{j}\frac{\partial\Phi}{\partial b_{j}}=n\, \Phi$.
Applying the Euler theorem to $\Phi$ we get
\begin{equation}
\sum_{i=1}^{r}a_{i}\frac{\partial\Phi}{\partial a_{i}}+
\sum_{j=1}^{n}b_{j}\frac{\partial\Phi}{\partial b_{j}}=
\sum_{i=1}^{r}a_{i}\frac{\partial\Phi}{\partial a_{i}}+ n\;\Phi=
\frac{n+r+1}{2}\Phi, \label{Euler1}
\end{equation}
thus
\begin{equation}\sum_{i=1}^{r}a_{i}\frac{\partial\Phi}{\partial
a_{i}}=\frac{r+1-n}{2}\Phi\neq 0, \label{Euler2}
\end{equation}
and from equation (\ref{COB}) we conclude that $\omega$ defines a
linear contact form on $\frak{g}$.

\medskip

Therefore the contact form is deeply related to the homogeneity
properties of the Casimir operator, but also that the extension
$\widehat{\frak{g}}$ used in \cite{C33}, which is derived from it.
This explains naturally why the determinant method developed there
holds. In particular, the contact form implies that these algebras
contract onto the Heisenberg algebra of the same dimension
\cite{C23}.

\subsection{The Lie algebras $\frak{sa}\left( N,\mathbb{R}\right) $}

The physically most important non-simple Lie algebra with only one
Casimir operator is the special affine algebra
$\frak{sa}(n,\mathbb{R})=\frak{sl}(n,\mathbb{R})\overrightarrow{\oplus}_{R}nL_{1}$
appearing in quantum gauge theories of gravity \cite{He,Pe}. The
invariant has been used both in the classification of particles,
as well as in establishing a wave equation \cite{Nee}. Various
works have been devoted to the problem of finding explicit
expressions of the Casimir operator of this algebra, either from
the analytical point of view \cite{He,Pe2}, with algebraic
procedures \cite{C33}, or more recently with the tensor approach
of enveloping algebras \cite{WW}. The advantage of the
Maurer-Cartan equations is the possibility of computing the
Casimir operator starting from an arbitrary basis. Taking for
example the boson realization of $\frak{sa}\left(
N,\mathbb{R}\right) $ given by
\begin{equation}
X_{\mu} =b_{\mu}^{+}b_{\mu}^{-}-b_{\mu+1}^{+}b_{\mu+1}^{-},\quad
X_{\mu\nu}   =b_{\mu}^{+}b_{\nu}^{-},\quad Y_{\nu} =b_{\nu}^{+},
\end{equation}
where $\left[  b_{i}^{+},b_{j}^{+}\right]  =\left[  b_{i}^{-},b_{j}%
^{-}\right]  =0,\;\left[  b_{i}^{-},b_{j}^{+}\right]
=\delta_{i}^{j}$, the corresponding Maurer-Cartan equations are
given by
\begin{eqnarray*}
d\omega_{\mu}  &
=\sum_{\rho=1}^{\mu}\sum_{\nu=\mu+1}^{N}\omega_{\rho\nu
}\wedge\omega_{\nu\rho},\;1\leq\mu\leq N-1\\
d\omega_{\nu\rho}  & =\sum_{\mu=1}^{N-1}\left(  \delta_{\nu}^{\mu}%
+\delta_{\rho}^{\mu+1}-\delta_{\rho}^{\nu}-\delta_{\nu}^{\mu+1}\right)
\omega_{\mu}\wedge\omega_{\nu\rho}+\sum_{\sigma=1}^{N}\omega_{\nu\sigma}%
\wedge\omega_{\sigma\rho},\\
d\theta_{\rho}  & =\delta_{\mu}^{\rho}\omega_{\mu}\wedge\theta_{\rho}%
-\delta_{\mu+1}^{\rho}\omega_{\mu}\wedge\theta_{\rho}+\delta_{\mu}^{\nu}%
\omega_{\rho\nu}\wedge\theta_{\mu}.
\end{eqnarray*}
Let $\alpha^{\mu},\beta^{\mu\nu},\gamma^{\rho}\in\mathbb{R}$ be
constants and
define the 1-form%
\[
\psi=\alpha^{\mu}\omega_{\mu}+\beta^{\nu\rho}\omega_{\nu\rho}+\gamma^{\sigma
}\theta_{\sigma}.
\]
Then \begin{equation} \fl
\psi\wedge\left(  \bigwedge^{\frac{1}{2}\left(  N^{2}+N-2\right)  }%
d\psi\right)  =\left(  \frac{N^{2}+N}{2}\right)  !\,C\left(
\alpha^{\mu
},\beta^{\nu\rho},\gamma^{\sigma}\right)  \left(  \bigwedge_{\mu=1}%
^{N-1}\omega_{\mu}\right)  \wedge\left(
\bigwedge_{\nu<\rho}\omega_{\nu\rho }\right)  \wedge\left(
\bigwedge_{\sigma=1}^{N}\theta_{\sigma}\right)  .
\end{equation}
Taking the polynomial $C\left(
\alpha^{\mu},\beta^{\nu\rho},\gamma^{\sigma }\right)  $ and
replacing the variables
\[
\alpha^{\mu}\longmapsto x_{\mu},\;\beta^{\nu\rho}\longmapsto
x_{\nu\rho },\;\gamma^{\sigma}\longmapsto p_{\sigma},
\]
the symmetrization of $C\left(
x_{\mu},x_{\nu\rho},p_{\sigma}\right)  $ provides the Casimir
operator of $\frak{sa}\left(  N,\mathbb{R}\right)$. So, for
example, for the lowest values of $N=2,3,4$, the previous function
$C\left( x_{\mu},x_{\nu\rho},p_{\sigma}\right)  $ has $3,58$ and
$8196$ terms, coinciding with the result of \cite{Pe2}.

\section{Algebras with a rational invariant}

The method, as presented, seems to be valid only for Lie algebras
having one classical Casimir operator. However, since rational
invariants of Lie algebras are the quotient of commuting
polynomials, it can be asked whether for the case of one rational
invariant, the commuting polynomials can be obtained from the
corresponding Maurer-Cartan equations.\footnote{Observe that in
this case, the condition $\frak{g}=
\left[\frak{g},\frak{g}\right]$ is no more satisfied \cite{AA}.}
This turns out to be the case, as we shall show with some
examples.

\subsection{The extended Poincar\'e algebra}

The Weyl group $W(1,3)$, i.e., the Poincar\'e group
$I\frak{so}(1,3)$ enlarged with a spacetime dilation operator $D$,
appears naturally when studying the symmetries of a spinning
particle which satisfies the Dirac equation when quantized
\cite{Ri}, constitutes an important example of a Lie algebra
exhibiting only one rational invariant. Using the standard basis
$E_{\mu\nu}=-E_{\nu\mu},P_{\mu}$ and metric $g={\rm diag}\left(
1,,1,1,-1\right)$, the brackets are given by
\begin{eqnarray}
\left[ E_{\mu \nu },E_{\lambda \sigma }\right]  =g_{\mu \lambda
}E_{\nu \sigma }+g_{\mu \sigma }E_{\lambda \nu }-g_{\nu \lambda
}E_{\mu \sigma }-g_{\nu \sigma }E_{\lambda \mu },\nonumber\\
\left[ E_{\mu \nu },P_{\rho }\right]  =g_{\mu \rho }P_{\nu
}-g_{\nu \rho }P_{\mu },\quad \left[ D,P_{\rho }\right]  =-P_{\rho
}.
\end{eqnarray}
Denoting by $\left\{\omega_{\mu\nu},\theta_{\mu},\xi \right\}$ the
dual basis, the Maurer-Cartan equations are simply
\begin{equation}
d\omega_{\nu \sigma}=g_{\mu \lambda}\omega_{\mu \nu}\wedge
\omega_{\lambda\sigma},\quad
d\theta_{\nu}=g_{\mu\rho}\omega_{\mu\nu}\wedge\theta_{\rho}-\xi\wedge\theta_{\nu},\quad
d\xi=0.
\end{equation}
A generic 1-form
$\psi=a^{\mu\nu}\omega_{mu\nu}+\beta^{\rho}\theta_{\rho}+\xi$ has
coboundary operator
$d\psi=a^{\mu\nu}d\omega_{mu\nu}+\beta^{\rho}d\theta_{\rho}$.
Since $\mathcal{N}(W(1,3))=1$, we obtain
\begin{equation}
\left(\bigwedge^{5}d\psi\right)\wedge\psi =
C_{1}\left(a^{\mu\nu},\beta^{\rho}\right)C_{2}\left(a^{\mu\nu},\beta^{\rho}\right)
\omega_{12}\wedge...\wedge\omega_{34}\wedge\theta_{1}\wedge...\wedge\theta_{4},
\end{equation}
where, after the corresponding replacements $\left\{
a^{\mu\nu}\mapsto e_{\mu\nu},\; b^{\rho}\mapsto p_{\rho}\right\}$,
the polynomials are given by $C_{1}=g^{\mu\mu}p_{\mu}^{2}$ and
\begin{eqnarray}
\fl C_{2}=-2\sum_{\mu<\nu<\rho}g_{\mu\mu}g_{\nu\nu}g_{\rho\rho}
\left(\epsilon_{\mu\nu\rho}p_{\mu}p_{\nu}e_{\mu\rho}e_{\nu\rho}
+\epsilon_{\mu\rho\nu}p_{\mu}p_{\rho}e_{\mu\nu}e_{\nu\rho}+
\epsilon_{\nu\rho\mu}p_{\nu}p_{\rho}e_{\nu\rho}e_{\mu\rho}\right)\nonumber\\
\lo
+\sum_{\mu<\nu}g_{\mu\mu}g_{\nu\nu}e_{\mu\nu}^{2}\left(\sum_{\rho\neq
\mu,\nu}g_{\rho\rho}p_{\rho}^{2}\right).
\end{eqnarray}
These are the well known Casimir operators of the Poincar\'e
algebra the quotient $C_{2}/C_{1}$ of which provides the invariant
of $W(1,3)$. Therefore the Maurer-Cartan equations provide the
commuting polynomials appearing in the rational invariant.

\subsection{The optical Lie algebra}

The optical Lie algebra $Opt(1,2)$ corresponds to the subalgebra
of the anti De Sitter algebra $\frak{so}(2,3)$ that leaves a
lightlike vector invariant in Minkowski spacetime \cite{Pa}. The
algebra $Opt(1,2)$ is defined by the non-vanishing brackets
\begin{equation}
\fl
\begin{array}
[c]{llll}%
\left[  K_{1},K_{2}\right]  =-K_{3}, & \left[  K_{1},K_{3}\right]
=-K_{2}, &
\left[  K_{2},K_{3}\right]  =K_{1}, & \left[  K_{1},M\right]  =-\frac{1}%
{2}M,\\
\left[  K_{1},Q\right]  =\frac{1}{2}Q, & \left[  K_{2},M\right]  =\frac{1}%
{2}Q, & \left[  K_{2},Q\right]  =\frac{1}{2}M, & \left[
K_{3},M\right]
=-\frac{1}{2}Q,\\
\left[  K_{3},Q\right]  =\frac{1}{2}M, & \left[  W,M\right]
=\frac{1}{2}M, & \left[  W,Q\right]  =\frac{1}{2}Q, & \left[
W,N\right]  =N.
\end{array}
\end{equation}
over the basis $\left\{  K_{i},W,M,Q,N\right\}  $. This algebra
has only one invariant, which is a rational function \cite{Pa}.
Again, the Maurer-Cartan equations provide information. Denoting
the dual basis by $\left\{
\omega_{1},\omega_{2},\omega_{3},\theta_{1},\theta_{2},\theta
_{3},\theta_{4}\right\}  $ and considering a linear combination
$\zeta
=k_{i}\omega_{i}+w\theta_{1}+m\theta_{2}+q\theta_{3}+n\theta_{4}$,
the result of the wedge product
\begin{equation*}
\fl \bigwedge^{3}d\zeta\wedge\zeta=3n\left(  q^{2}\left(
k_{2}+k_{3}\right) +m^{2}\left(  k_{3}-k_{2}\right)
-2k_{1}mq\right)  \omega_{1}\wedge
..\omega_{3}\wedge\theta_{1}\wedge..\wedge\theta_{4}
\end{equation*}
shows that $Opt(1,2)$ is endowed with a contact form. Although the
polynomial $C=3n\left( q^{2}\left( k_{2}+k_{3}\right) +m^{2}\left(
k_{3}-k_{2}\right) -2k_{1}mq\right)$ in not an invariant of
$Opt(1,2)$, it can be easily verified that any term is a
semi-invariant, from which the rational invariant is easily
deduced as the quotient of the terms $P=\left( q^{2}\left(
k_{2}+k_{3}\right) +m^{2}\left( k_{3}-k_{2}\right)
-2k_{1}mq\right)/n$.

\subsection{The two-photon algebra}
The 2-photon algebra $\frak{h}_{6}$, isomorphic to the
Schr\"odinger algebra in $(1+1)$-dimension, has been used, among
other applications, to construct infinite classes of $N$-particle
Hamiltonian systems \cite{Zha,Her}. Over the basis $\left\{
N,A_{+},A_{-},B_{+},B_{-},M\right\}  $ the brackets are given by
\[%
\begin{array}
[c]{lll}%
\left[  N,B_{\pm}\right]  =\pm2B_{\pm}, & \left[  N,A_{\pm}\right]
=\pm
A_{\pm}, & \left[  B_{+},B_{-}\right]  =-4N-2M,\\
\left[  B_{+},A_{-}\right]  =-2A_{+}, & \left[  B_{-},A_{+}\right]
=2A_{-}, & \left[  A_{+},A_{-}\right]  =-M.
\end{array}
\]
This algebra clearly has two Casimir operators, one being the
central charge $M$. To compute them using Maurer-Cartan equations,
we consider the extension of $\frak{h}_{6}$ by an element $Y$
acting on the two-photon algebra as
follows:%
\[
\left[  Y,N\right]  =-M,\;\left[  Y,A_{\pm}\right]
=A_{\pm},\;\left[ Y,M\right]  =2M.
\]
The seven dimensional algebra $\frak{g}$ satisfies
$\mathcal{N}\left( \frak{g}\right)  =1$. Let $\left\{
\vartheta,\omega_{1},\omega_{2},\theta
_{1},\theta_{2},\sigma,\chi\right\}  $ be the dual basis to
$\left\{
N,A_{+},A_{-},B_{+},B_{-},M,Y\right\}  $ and take the structure equations%
\[%
\begin{tabular}
[c]{l}%
$d\vartheta=-4\theta_{1}\wedge\theta_{2},\;d\theta_{1}=2\vartheta\wedge
\theta_{1},\;d\theta_{2}=-2\vartheta\wedge\theta_{2},\;d\chi=0,$\\
$d\omega_{1}=\vartheta\wedge\omega_{1}+2\omega_{2}\wedge\theta_{1}+\omega
_{1}\wedge\chi,$\\
$d\omega_{2}=-\vartheta\wedge\omega_{2}-2\omega_{1}\wedge\theta_{2}+\omega
_{2}\wedge\chi,$\\
$d\sigma=-2\theta_{1}\wedge\theta_{2}-\omega_{1}\wedge\omega_{2}%
-\vartheta\wedge\chi+2\sigma\wedge\chi.$%
\end{tabular}
\]
An arbitrary linear combination $\alpha=a_{1}\omega_{1}+a_{2}\omega_{2}%
+a_{3}\vartheta+a_{4}\theta_{1}+a_{5}\theta_{2}+a_{6}\sigma+a_{7}\chi$
gives
rise to the wedge product%
\begin{eqnarray*}
\fl \alpha\wedge d\alpha\wedge d\alpha\wedge d\alpha =& 12a_{6}\left(  a_{6}{}%
^{3}-4a_{6}\left(  a_{4}a_{5}+a_{1}a_{2}\right)  +a_{3}a_{6}\left(
a_{3}+a_{6}\right)
+a_{1}^{2}a_{5}+a_{4}a_{2}^{2}-2a_{1}a_{2}a_{3}\right)
\\
 &\times
\omega_{1}\wedge\omega_{2}\wedge\vartheta\wedge\theta_{1}\wedge\theta
_{2}\wedge\sigma\wedge\chi.
\end{eqnarray*}
With the replacements%
\[%
\begin{array}
[c]{cccccc}%
a_{1}\mapsto a_{+}, & a_{2}\mapsto a_{-}, & a_{3}\mapsto n, &
a_{4}\mapsto b_{+}, & a_{5}\mapsto b_{-}, & a_{6}\mapsto m,
\end{array}
\]
we obtain the polynomial%
\begin{equation}
\fl P=12m\left(  m^{3}-4m\left(  b_{+}b_{-}+a_{+}a_{-}\right)
+mn\left( n+m\right)
+a_{+}^{2}b_{-}+b_{+}a_{-}^{2}-2a_{+}a_{-}n\right).
\end{equation}
$P$ is the Casimir operator of the extended algebra $\frak{g}$,
and it is straightforward to verify that $C_{1}=m$ and
$C_{2}=-4m\left(
b_{+}b_{-}+a_{+}a_{-}\right)  +mn\left(  n+m\right)  +a_{+}^{2}b_{-}%
+b_{+}a_{-}^{2}-2a_{+}a_{-}n$ are the invariants of the subalgebra
$\frak{h}_{6}$.

\bigskip Although in these examples the Maurer-Cartan
equations do not provide the rational invariant itself, they give
the commuting polynomials that intervene in it.\footnote{More
specifically, they provide the Casimir operators of a codimension
one subalgebra that is perfect.} This means that the invariant has
the form $C=C_{1}^{a}C_{2}^{-b}$, where only the values of $a,b$
have to be checked. An interesting observation concerning these
and other examples is that the degree of the product of these
polynomials is always the integer part of half the dimension of
the algebra plus one. Analyzing with the same method all Lie
algebras of odd dimension $n\leq 9$ with one rational invariant
and non-trivial Levi decomposition \cite{Zas,Pa2}, we found always
the same rule for the degrees of $C_{1}$ and $C_{2}$. This enables
us to suggest the following general property for Lie algebras:

\begin{lemma}
Let $\frak{g}=\frak{s}\overrightarrow{\oplus}_{R}\frak{r}$ be a
Lie algebra satisfying the constraint $\dim \left(\frak{g}/\left[
\frak{g},\frak{g}\right]\right)\leq 1$ and
$\mathcal{N}(\frak{g})=1$. Then the rational invariant
$C=C_{1}^{a}C_{2}^{-b}$ is such that
\begin{equation}
\deg \left(C_{1} C_{2}\right)\leq
\frac{1}{2}\left(\dim\frak{g}+1\right).
\end{equation}
\end{lemma}

The previous examples correspond to the case $\dim\left(
\frak{g}/\left[ \frak{g},\frak{g}\right]  \right) =1$, where
$\frak{g}^{\prime}=\left[  \frak{g,g}\right]  $ is a codimension
one subalgebra that satisfies $\frak{g}^{\prime}=\left[
\frak{g}^{\prime},\frak{g}^{\prime}\right]  $, thus admits Casimir
operators as invariants. Since $\mathcal{N}\left( \frak{g}\right)
=1$, this means that $\mathcal{N}\left( \frak{g}^{\prime}\right)
=2$, and therefore the polynomials $C_{1}$ and $C_{2}$ found for
the rational invariant of $\frak{g}$ can be taken as the Casimir
operators of $\frak{g}^{\prime}$.

\section{Final remarks}

It has been proved that for Lie algebras
$\frak{g}=\frak{s}\overrightarrow{\oplus}_{R}nL_{1}$ with one
Casimir operator, the latter can be obtained by means of wedge
products of the Maurer-Cartan equations. This constitutes a
generalization, to non-simple Lie algebras, of the well known fact
that for the simple algebra $\frak{su}(2)$ the quadratic Casimir
operator arises from the structure equations. This fact shows
moreover that for this class of algebras, the invariant inherits a
clear geometrical meaning as the function that appears in the
volume form determined by the structure equations. The proof bases
on the homogeneity properties of the Casimir operators of
inhomogeneous Lie algebras with respect to the variables of the
Levi part and the representation describing the semidirect
product. In particular, these property allows to obtain sharp
bounds for the dimension of the representation space $R$, whenever
the constraint $\mathcal{N}(\frak{g})=1$ is satisfied. This in
principle enables us to classify inhomogeneous algebras with only
one invariant, given a fixed Levi subalgebra. Further, it has been
pointed out that the procedure based on differential forms can be
extended for Lie algebras $\frak{g}$ having only one rational
invariant and a codimension one commutator subalgebra. This
enables us further to reconstruct the Casimir operators of the
latter as the terms appearing in the volume form associated to
$\frak{g}$, thus providing a geometrical method to compute the
Casimir operators of Lie algebras with two invariants.

\medskip

The procedure using differential forms is clearly limited to Lie
algebras having rational invariants. Since the function appearing
in the volume form obtained from the Maurer-Cartan equations is
always polynomial, Lie algebras having transcendental invariants,
like most solvable Lie algebras, are excluded. Whether introducing
additional constraints to the ansatz of forms can provide a method
to cover solvable algebras with a non-rational invariant is still
an open question. Nowadays, the best possible known approach for
the solvable case, using forms, is that of the moving frame method
developed in \cite{Bo,Bo2}.

\medskip

Finally, the differential-geometric derivation of the Casimir
operator developed here serves to clarify a result concerning the
degrees of a Casimir operator. In \cite{Pe}, it was claimed that
for an $n$-dimensional Lie algebra $\frak{g}$ with one invariant,
the corresponding Casimir operator has either degree one or
$\frac{n+1}{2}$. We give a counterexample to this claim, showing
that the degree can be actually different. Consider the Lie
algebra $L_{7,2}$ determined by the Maurer-Cartan equations
\begin{equation*}
\fl\begin{tabular}
[c]{ll}%
\multicolumn{2}{l}{$d\omega_{1}=\omega_{2}\wedge\omega_{3},\;d\omega
_{2}=-\omega_{1}\wedge\omega_{3},\;d\omega_{3}=\omega_{1}\wedge\omega_{2},$}\\
$d\omega_{4}=-\frac{1}{2}\left(  \omega_{1}\wedge\omega_{7}-\omega_{2}%
\wedge\omega_{5}+\omega_{3}\wedge\omega_{6}\right)  ,$ & $d\omega_{5}%
=-\frac{1}{2}\left(
\omega_{1}\wedge\omega_{6}-\omega_{2}\wedge\omega
_{4}+\omega_{3}\wedge\omega_{5}\right)  $,\\
$d\omega_{6}=\frac{1}{2}\left(  \omega_{1}\wedge\omega_{5}-\omega_{2}%
\wedge\omega_{7}+\omega_{3}\wedge\omega_{4}\right)  ,$ & $d\omega_{7}%
=-\frac{1}{2}\left(
\omega_{1}\wedge\omega_{4}+\omega_{2}\wedge\omega
_{6}-\omega_{3}\wedge\omega_{5}\right)  $.%
\end{tabular}
\end{equation*}
Since the algebra is indecomposable, it  has no invariant of order
one. Following theorem 1 of \cite{Pe}, the Casimir operator should
be of order four. However, taking a generic element
$\omega=a_{i}\omega_{i}$ and computing the corresponding wedge
product $\omega\wedge \left(\bigwedge^{3} d\omega\right)$ we
obtain the expression
\begin{equation*}
\omega\wedge \left(\bigwedge^{3}
d\omega\right)=(a_{4}^{2}+a_{5}^{2}+a_{6}^{2}+a_{7}^{2})^{2}\omega_{1}\wedge..\wedge\omega_{7},
\end{equation*}
and the function $(a_{4}^{2}+a_{5}^{2}+a_{6}^{2}+a_{7}^{2})^{2}$
is a square. It can be easily verified that
$x_{4}^{2}+x_{5}^{2}+x_{6}^{2}+x_{7}^{2}$ is a quadratic Casimir
invariant of the algebra. The reason for the failure of the
statement in \cite{Pe} lies in the fact that the independence of
the invariant on the variables of the Levi part was not explicitly
considered in the proof. In this sense, the result of \cite{Pe}
should be reformulated as

\begin{proposition}
If $\frak{g}=\left[\frak{g},\frak{g}\right]$ has only one
invariant $C$, then the order of $C$ is either one or a divisor of
$\frac{1}{2}\left(\dim\frak{g}+1\right)$.
\end{proposition}

\section*{Acknowledgment}
During the preparation of this work, the author was financially
supported by the research project MTM2006-09152 of the M.E.C. and
the project and CCG07-UCM/ESP-2922 of the U.C.M.-C.A.M.

\end{document}